\documentclass[letterpaper,journal]{IEEEtran}
\usepackage[T1]{fontenc}
\usepackage{amsmath,amssymb,amsfonts,amsthm}
\usepackage{graphicx,xcolor,booktabs,array,tabularx,enumerate}
\usepackage{cite,url}
\allowdisplaybreaks[1]
\newtheorem{theorem}{Theorem}
\newtheorem{lemma}{Lemma}

\usepackage[hidelinks]{hyperref}
\hypersetup{pdftitle={Two- and three-population evolutionary games with random payoffs},pdfauthor={Z. Song, M. H. Duong, T. A. Han}}
\graphicspath{{figures/}}
\usepackage{comment}
\newcommand{\hide}[1]{}

\title{Equilibrium Statistics and Stability in Random Two-Population and Three-Population  Evolutionary Games}

\author{Zhao~Song, Manh~Hong~Duong, and The~Anh~Han%
\thanks{\textit{(Corresponding author: Zhao Song.)}}
\thanks{Z. Song and T. A. Han are with the School of Computing, Engineering and Digital Technologies, Teesside University, UK (e-mail: songzhao.songz@gmail.com; t.han@tees.ac.uk).}%
\thanks{M. H. Duong is with the School of Mathematics,
University of Birmingham, UK (e-mail: h.duong@bham.ac.uk).}%
}

\begin{document}
\maketitle
\begin{abstract}
Multi-population matrix evolutionary games provide a framework for studying adaptation and strategic interactions among heterogeneous agents. When payoffs are uncertain, however, analysing a specified game does not reveal how frequently different evolutionary outcomes occur or how their probabilities depend on modelling assumptions. We study two- and three-population games with two strategies per population and both within- and between-population pairwise interactions, with payoffs fixed during each evolutionary trajectory. For two populations, we derive equilibrium-existence probabilities and analytical stability results for generic games and three structured classes: diversity, exclusive, and avoidance games. For generic two-population games, independent Gaussian sampling of payoff entries and of replicator coefficients produces different equilibrium statistics. Interaction structure further distinguishes coexistence from attraction: diversity and exclusive games share the same interior-existence probability, but only exclusive games admit asymptotically stable interior equilibria. Avoidance games admit transversely attracting equilibrium families without locally asymptotically stable individual points. For generic three-population games under independent payoff sampling, we obtain existence probabilities for all equilibrium types and joint existence-and-stability probabilities for boundary equilibria. Monte Carlo simulations support the analytical predictions and quantify multistability, while phase portraits illustrate initial-condition dependence. Under the same Gaussian payoff-sampling protocol, simulations show a lower frequency of fully mixed equilibria but a higher frequency of multistability for three populations than for two. These results provide probabilistic benchmarks for distinguishing coexistence from attraction and assessing how uncertainty representation, interaction structure, and population number shape evolutionary outcomes.
\end{abstract}
\begin{IEEEkeywords}
Evolutionary game theory, random payoffs, multi-population games, replicator dynamics, equilibrium stability, interaction structure.
\end{IEEEkeywords}

\section{Introduction}

Collective adaptation in heterogeneous multi-agent systems depends on both the incentives facing individual groups and the interactions that connect them. The importance of coupled populations is well established in ecological dynamics~\cite{lotka1926elements,volterra1926fluctuations,gavina2018multi}. In strategic settings, comparable interdependence arises among governments, enterprises, and communities in resource provision and environmental governance~\cite{zhang2022tripartite,zhao2020tripartite,zhou2022evolutionary,encarnaccao2016paradigm}, and among developers, users, regulators, and media in AI governance~\cite{alalawi2026trust,balabanova2025media,da2025can}. Multi-population evolutionary games describe adaptation among groups with different roles~\cite{taylor1979evolutionarily}, while networked extensions highlight the effects of within- and between-group interactions~\cite{liu2024heterogeneously,vargas2026modeling,cimpeanu2023social,vargas2026out}. Evolutionary dynamics also connects collective-behaviour modelling with computational applications such as sensor identification~\cite{tan2016evolutionary,yazidi2020solving}. When interaction costs and benefits are uncertain, however, analysing one specified game cannot establish how representative its outcomes are~\cite{duong2025evolutionary}.

Existing research shows that evolutionary outcomes depend on how interactions and adaptation jointly determine the relative success of competing strategies. Even with an unchanged strategy set, differences in interaction structure can change whether cooperation and fairness persist~\cite{jia2025social,jia2025asymmetric,wei2025indirect,guo2023third,song2026evolution}. Persistence, however, is not synonymous with convergence: coexistence can occur at an attracting mixed equilibrium or through sustained oscillations, while multiple attractors can make the realised outcome depend on initial conditions~\cite{11743107,wang2023emergence,bauer2019stabilization,ijcai2022p76,galla2013complex}. Analyses of asymmetric games further show that equilibria need not be isolated: continuous equilibrium sets require their own stability analysis~\cite{song2024study,song2024stability}. Alongside these dynamical insights, random-game theory quantifies equilibrium occurrence across uncertain payoff realisations~\cite{galla2007two,han2012equilibrium,duong2021statistics,duong2025random}. Connections between equilibrium equations and random polynomial systems make it possible to characterise the expected number of interior equilibria and the distribution of their counts~\cite{gokhale2010evolutionary,duong2016analysis,duong2018distribution}. Together, these advances distinguish three questions: which equilibria are possible, how frequently they occur, and whether they attract the dynamics.

However, connecting these questions for populations with both within- and between-population interactions requires a consistent treatment of uncertainty, interaction structure, and stability. Interior-equilibrium counts alone do not describe boundary outcomes, and coexistence does not necessarily imply attraction. Moreover, specifying uncertainty in payoff entries is not equivalent to specifying it directly in replicator coefficients: shared payoff entries induce dependence between the coefficients, even when the entries themselves are independent. Restrictions on which interactions contribute to payoffs can further change both equilibrium frequencies and whether equilibria are isolated or form continuous families. These issues concern variation across fixed games, distinct from stochastic forcing in population dynamics or changes of game during evolution~\cite{hong2026multi,feng2022stochastic,feng2023evolutionary}. Addressing them within a common framework would provide benchmarks when precise payoffs are unavailable and clarify which conclusions depend on the assumed representation of uncertainty, rather than on interaction structure alone.

To address these questions, we study generalised evolutionary games with two strategies per population under Gaussian uncertainty, focusing on two and three populations. For generic two-population games, we compare independent payoff sampling, which starts from the game matrix, with independent coefficient sampling, which starts from the replicator equations but omits the dependence induced by shared payoffs. We then impose zero-block constraints to obtain diversity, exclusive, and avoidance games, sampling only the active payoff entries. Gaussian probability calculations and the Kac--Rice formula~\cite{azais2009level} yield equilibrium-existence probabilities and analytical stability results. For three populations, we retain pairwise interactions and derive existence probabilities for all candidate types and joint existence-and-stability probabilities for boundary candidates; interior stability is assessed numerically. Monte Carlo simulations test the predictions and quantify multistability, while phase portraits illustrate initial-condition dependence and non-equilibrium attraction. Payoffs remain fixed during each trajectory, so the randomness describes variation across games rather than temporal noise.

In summary, the main contributions of this paper are as follows: 
\begin{enumerate} 
\item We derive equilibrium-existence and joint existence-and-stability probabilities for generic two-population games under payoff and coefficient sampling, showing how uncertainty representation changes equilibrium statistics. 
\item We determine how three interaction structures change equilibrium occurrence and stability: diversity games have no asymptotically stable interior equilibrium, exclusive games can support stable interior coexistence, and avoidance games admit transversely attracting equilibrium families without asymptotically stable individual points. 
\item We extend the generic analysis to three populations under payoff sampling, deriving existence probabilities for all candidate types and joint existence-and-stability probabilities for boundary candidates. Simulations quantify multistability and three-population interior stability, while two-population phase portraits illustrate initial-condition dependence. 
\end{enumerate}

Section~\ref{sec: multi-population EGT} defines the models and numerical protocol. Sections~\ref{sec: 2-population} and ~\ref{sec:3pop-analysis} present two- and three-population results. Section~\ref{sec: numerics} reports numerical results and Section~\ref{conclusion} concludes. Proofs, additional simulations, and model extensions are provided in the Supplementary Material.

\section{Models and Methods}
\label{sec: multi-population EGT}
\subsection{General setting}
Consider $m$ infinitely large, well-mixed populations with strategy sets $S_j=\{1,\ldots,n_j\}$ \cite{taylor1979evolutionarily,bauer2019stabilization}. Let $x_{ij}$ be the frequency of strategy $i$ in population $j$, with $\sum_{i\in S_j}x_{ij}=1$, and write $x_j=(x_{ij})_{i=1}^{n_j}$ and $x=(x_1,\ldots,x_m)$. If $\pi_{ij}(x)$ is the strategy payoff, the population average payoff and replicator dynamics are
\begin{equation}
\bar{\pi}_j(x)=\sum_{i\in S_j} x_{ij}\pi_{ij}(x),
\label{eq2}
\end{equation}
\begin{equation}
\dot{x}_{ij}=x_{ij} (\pi_{ij}(x)-\bar{\pi}_{j}(x)), \quad 1\leq j\leq m,\quad i\in S_j. 
\label{eq3}
\end{equation}
Thus, a strategy increases in frequency when its payoff exceeds its population average. The frequency constraints reduce the system to $\sum_{j=1}^m(n_j-1)$ independent coordinates.

When each population has two strategies, label them $s_{2i-1}$ and $s_{2i}$, and let $x_i$ denote the frequency of the former. With $x=(x_1, \ldots, x_m)$ and $i=1, \ldots, m$,
\begin{equation}
f_i(x):=\pi_{s_{2i-1}}(x)-\pi_{s_{2i}}(x),
\label{eq: payoff-difference}
\end{equation}
\begin{equation}
\label{eq: RE2}
\begin{gathered}
  \dot{x}_i=x_i(1-x_i)f_i(x).
\end{gathered}
\end{equation}
Each encounter involves two players from the same or different populations. Let $\beta_{rs}$ be the payoff to strategy $s_r$ against $s_s$. With equal interaction weights, the expected payoff for $r=1, \ldots, 2m$ is
\begin{equation}
\label{eq: pairwise-payoff}
\begin{gathered}
\pi_{s_r}(x)=\sum_{j=1}^{m} \bigl[\beta_{r,2j-1}x_j+\beta_{r,2j}(1-x_j)\bigr].
\end{gathered}
\end{equation}
Averaging over populations instead would multiply all payoffs by $1/m$ and only rescale time.

A state $x^*\in[0,1]^m$ is an equilibrium if and only if
\begin{equation}
x_i^*(1-x_i^*)f_i(x^*)=0,\qquad i=1,\ldots,m.
\label{eq: equilibrium-condition}
\end{equation}
Interior equilibria satisfy $f_i(x^*)=0$ for every $i$; boundary equilibria require the full condition (\ref{eq: equilibrium-condition}). Pure coordinates represent one strategy, while mixed coordinates represent coexistence of both strategies in that population.

Unless otherwise specified, stability below means local asymptotic stability relative to the feasible state space: nearby feasible states remain nearby and converge to the equilibrium.

\subsection{Two-population two-player two-strategy games}
\label{222}
For $m=2$, the strategy sets are $\{s_1,s_2\}$ and $\{s_3,s_4\}$. The diagonal and off-diagonal blocks of $B=(\beta_{rs})$ describe within- and between-population payoffs, respectively:
\begin{equation}
\bordermatrix{
       & s_1        & s_2       & s_3    &   s_4 \cr
s_1     &  \beta_{11} & \beta_{12} & \beta_{13} & \beta_{14} \cr
s_2     & \beta_{21} & \beta_{22} & \beta_{23} & \beta_{24} \cr
s_3    & \beta_{31} & \beta_{32} & \beta_{33} & \beta_{34} \cr
s_4     & \beta_{41} & \beta_{42} & \beta_{43} & \beta_{44}
},
\end{equation}


Write $x=x_1$ and $y=x_2$ for the frequencies of $s_1$ and $s_3$, respectively. Substituting \eqref{eq: pairwise-payoff} into \eqref{eq: payoff-difference} gives
\begin{align*}
f_1(x,y)&=a_1x+a_2y+a_3,\\
f_2(x,y)&=b_1x+b_2y+b_3,
\end{align*}
where
\begin{align*}
a_1&=\beta_{11}-\beta_{21}+\beta_{22}-\beta_{12},\\
a_2&=\beta_{13}-\beta_{23}+\beta_{24}-\beta_{14},\\
a_3&=\beta_{12}-\beta_{22}+\beta_{14}-\beta_{24},\\
b_1&=\beta_{31}-\beta_{41}+\beta_{42}-\beta_{32},\\
b_2&=\beta_{33}-\beta_{43}+\beta_{44}-\beta_{34},\\
b_3&=\beta_{32}-\beta_{42}+\beta_{34}-\beta_{44}.
\end{align*}
Thus \eqref{eq: RE2} becomes
\begin{equation}
\begin{aligned}
\dot{x}&=x(1-x)f_1(x,y),\\
\dot{y}&=y(1-y)f_2(x,y).
\end{aligned}
\label{stystem}
\end{equation}

\subsection{Three-population two-player two-strategy games}
\label{sec:3pop-model}
For $m=3$, let $x, y, z$ be the frequencies of $s_1, s_3, s_5$ in the strategy pairs $\{s_1,s_2\}$, $\{s_3,s_4\}$, and $\{s_5,s_6\}$. All blocks of $B\in\mathbb R^{6\times6}$ are retained, with payoffs given by \eqref{eq: pairwise-payoff}.

Let $\boldsymbol u=(x,y,z)^T$ and write
\begin{equation}
\label{eq:3pop-affine}
\boldsymbol f(\boldsymbol u)=C\boldsymbol u+\boldsymbol d,\qquad
C=\begin{pmatrix}a_1&a_2&a_3\\b_1&b_2&b_3\\c_1&c_2&c_3\end{pmatrix},\quad
\boldsymbol d=\begin{pmatrix}a_4\\b_4\\c_4\end{pmatrix}.
\end{equation}
The payoff-to-coefficient mapping can be expressed compactly as
\begin{equation}
\label{eq:3pop-beta-map}
\begin{aligned}
C_{ij}&=\beta_{2i-1,2j-1}-\beta_{2i,2j-1}
          -\beta_{2i-1,2j}+\beta_{2i,2j},\\
d_i&=\sum_{j=1}^{3}(\beta_{2i-1,2j}-\beta_{2i,2j}),
\qquad i,j\in\{1,2,3\}.
\end{aligned}
\end{equation}
Thus the replicator dynamics on $[0,1]^3$ is
\begin{equation}
\label{3-RE}
\begin{aligned}
\dot x&=x(1-x)(a_1x+a_2y+a_3z+a_4),\\
\dot y&=y(1-y)(b_1x+b_2y+b_3z+b_4),\\
\dot z&=z(1-z)(c_1x+c_2y+c_3z+c_4).
\end{aligned}
\end{equation}
Deterministic three-party games with two strategies per party have also been analysed in \cite{song2024study,song2024stability}. Unlike those models, the present extension retains pairwise interactions, so the payoff differences remain affine in the population frequencies.

\subsection{Random payoffs}
The replicator dynamics depend on the payoff entries only through the coefficients of the payoff differences. Uncertainty can therefore be specified either in the underlying payoffs or directly in these coefficients. We distinguish two Gaussian sampling models:

\begin{itemize}
    \item \textit{Payoff sampling (microscopic uncertainty):} active payoff entries are sampled independently from $\mathcal N(0,1)$, and the replicator coefficients are then calculated from them. For structured games, excluded payoff blocks are fixed at zero.

    \item \textit{Coefficient sampling (macroscopic uncertainty):} the coefficients of the generic replicator equations are sampled independently from $\mathcal N(0,1)$.
\end{itemize}

Derived coefficients are generally correlated and need not have equal variance; independent coefficient sampling does not preserve this dependence. Sampled parameters remain fixed throughout each game's dynamics.

\subsection{Numerical protocol}
\label{sec:numerical-protocol}

Monte Carlo simulations use $M=10^6$ independent game realisations per setting under the sampling assumptions above. Comparisons across interaction structures share the same base payoff draws. Equilibrium candidates are computed algebraically and retained when all free coordinates lie strictly in $(0,1)$.

For isolated candidates, numerical local asymptotic stability requires every eigenvalue of the full replicator Jacobian to satisfy
\[
\operatorname{Re}\lambda<-\tau,
\qquad
\tau=10^{-12}+10^{-10}\max(1,\|J\|_{\infty}).
\]
Real parts within $[-\tau,\tau]$ are flagged for separate assessment. Structurally nonhyperbolic cases are classified using the analytical results.

All reported frequencies use the full $M$ realisations as their denominator. For candidate $i$, $E_i$ denotes existence and $S_i$ local asymptotic stability; thus $\widehat{\mathbb P}(E_i\cap S_i)$ is a joint probability. The counts $N$ and $N_{\mathrm s}$ denote existing and locally asymptotically stable equilibria. For avoidance games, they instead count equilibrium families and transversely attracting families, with transverse attraction denoted by $T$. Grouped entries report mean per-candidate frequencies; where shown, SD denotes the between-candidate standard deviation. Implementation details are provided in the Supplementary Material.

\section{Two-population analytical results}
\label{sec: 2-population}
We analyse generic games under both sampling models and diversity, exclusive, and avoidance games under independent Gaussian sampling of their active payoffs. Detailed proofs and expanded special-case formulas are provided in the Supplementary Material.

\subsection{Generic games}
\label{generic games}
In the nondegenerate case, there are four vertex candidates, four open-edge candidates, and one interior candidate. The latter five require their free coordinates to lie in $(0,1)$.
\begin{lemma}
\label{lem: possible equlibria}
Assume that the denominators below are nonzero, as holds almost surely under either generic sampling model. The equilibrium candidates of \eqref{stystem} and the corresponding Jacobian eigenvalues are as follows. An open-edge or interior candidate is an equilibrium only when all its free coordinates lie strictly in $(0,1)$.
\begin{itemize}
    \item at $A_1=(0,0)$: $\lambda_{1}=a_3$ and $\lambda_2=b_3$. 
    \item at $A_2=(0,1)$: $\lambda_1=a_2+a_3$ and $\lambda_2=-(b_2+b_3)$.
    \item at $A_3=(1,0)$: $\lambda_1=-(a_1+a_3)$ and $\lambda_2=b_1+b_3$.
    \item at $A_4=(1,1)$: $\lambda_1=-(a_1+a_2+a_3)$ and $\lambda_2=-(b_1+b_2+b_3)$.
    \item at $A_5=(0,-\frac{b_3}{b_2})$: $\lambda_1=-\frac{b_3(b_2+b_3)}{b_2}$ and $\lambda_2=\frac{a_3b_2-a_2b_3}{b_2}$.
    \item at $A_6=(1,-\frac{b_1+b_3}{b_2})$: $\lambda_1=-\frac{(b_1+b_3)(b_1+b_2+b_3)}{b_2}$, $\lambda_2=\frac{a_2b_1-a_1b_2-a_3b_2+a_2b_3}{b_2}$.
    \item at $A_7=(-\frac{a_3}{a_1},0)$: $\lambda_1=-\frac{a_3(a_1+a_3)}{a_1}$ and $\lambda_2=\frac{-a_3b_1+a_1b_3}{a_1}$.
    \item at $A_8=(-\frac{a_2+a_3}{a_1},1)$: $\lambda_1=-\frac{(a_2+a_3)(a_1+a_2+a_3)}{a_1}$ and $\lambda_2=\frac{a_2b_1+a_3b_1-a_1b_2-a_1b_3}{a_1}$.
    \item at $A_9=(x^*,y^*)=(\frac{a_2b_3-a_3b_2}{a_1b_2-a_2b_1},\frac{a_3b_1-a_1b_3}{a_1b_2-a_2b_1})$, $\lambda_1=\frac{-b+\sqrt{\Delta}}{2}$, $\lambda_2=\frac{-b-\sqrt{\Delta}}{2}$,
    where $b=-[a_1 x^*(1-x^*)+b_2 y^*(1-y^*)]$, $\Delta=[a_1 x^*(1-x^*)-b_2 y^*(1-y^*)]^2+4a_2b_1x^*(1-x^*) y^*(1-y^*)$.
\end{itemize}
\end{lemma}

The following theorem compares per-candidate existence and joint existence-and-stability probabilities under the two sampling models.

\begin{theorem}[Equilibrium analysis of generic two-population two-player two-strategy random games]\
\label{thm: main-thm}
\begin{enumerate}
    \item[(1)] \textit{Coefficient sampling (macroscopic uncertainty).}
    Suppose that $a_1,a_2,a_3,b_1,b_2,b_3$ are independent $\mathcal N(0,1)$ random variables. Each vertex $A_1,\ldots, A_4$ always exists and is locally asymptotically stable with probability $1/4$. Each of $A_5$ and $A_7$ exists with probability $1/4$ and exists and is locally asymptotically stable with probability $1/16$. For each of $A_6$ and $A_8$, the corresponding probabilities are
    \[
    p_{\mathrm{edge}}
    =\frac{1}{\pi}\arctan\frac{1}{\sqrt{2}}
    \approx0.1959,
    \qquad
    \frac{p_{\mathrm{edge}}}{4}\approx0.0490.
    \] 
    The interior candidate $A_9$ exists with probability $1/12$ and exists and is locally asymptotically stable with probability $1/48$.
    
    \item[(2)] \textit{Payoff sampling (microscopic uncertainty).}
    Suppose that all payoff entries $\beta_{ij}$ are independent $\mathcal N(0,1)$ random variables. Each vertex $A_1,\ldots, A_4$ always exists and is locally asymptotically stable with probability $1/4$. Each open-edge candidate $A_5,\ldots, A_8$ exists with probability $1/3$ and exists and is locally asymptotically stable with probability $1/12$. The interior candidate $A_9$ exists with probability
    \[
    p_{\mathrm{int}}=\frac{2\arctan(\sqrt{2})-\arctan(2\sqrt{2})}{\pi}\approx0.21635,
    \]
    and exists and is locally asymptotically stable with probability $p_{\mathrm{int}}/4\approx0.05409$.
\end{enumerate}
\end{theorem}
Under either sampling model, the conditional probability of local asymptotic stability of the interior equilibrium, given its existence, is $1/4$. Numerical results below test these predictions and quantify the distributions of equilibrium counts within individual games. 
\textbf{Idea of proof:}
The calculations separate the candidates according to their location. Vertices always exist, so their stability probabilities follow from the signs of the two Jacobian eigenvalues. For open-edge candidates $A_5,\ldots, A_8$, existence requires the free coordinate to lie in $(0,1)$, and stability additionally requires both eigenvalues to be negative. These events are evaluated using the relevant joint Gaussian distributions, retaining the dependence induced by the chosen sampling model. For the interior candidate $A_9$, the Kac--Rice formula~\cite{azais2009level} gives its existence probability through the expected number of zeros of the affine payoff-difference system. Independent sign changes of the two payoff-difference equations yield the interior-stability probabilities. Detailed proofs are provided in the Supplementary Material.

The Supplementary Material further examines how changing the relative weights of within- and between-population interactions affects equilibrium statistics in a scaled payoff extension of the generic model.

We now keep the payoff-sampling assumption fixed and vary which interactions contribute to payoffs. Diversity games retain only between-population contributions, exclusive games retain only within-population contributions, and avoidance games retain contributions only within population 1. These restrictions isolate the roles of the two interaction types and reveal how they affect both equilibrium occurrence and attraction.
\subsection{Diversity games}
Diversity games retain only between-population payoffs: both diagonal blocks of $B$ are zero, giving $a_1 = b_2 = 0$. This structure includes two-strategy representations of role-separated games such as ultimatum, trust, and dictator games:
\begin{equation}
\bordermatrix{
       & s_1        & s_2       & s_3    &   s_4 \cr
s_1     &  0 & 0 & \beta_{13} & \beta_{14} \cr
s_2     & 0 & 0 & \beta_{23} & \beta_{24} \cr
s_3    & \beta_{31} & \beta_{32} & 0 & 0 \cr
s_4     & \beta_{41} & \beta_{42} & 0 & 0
},
\end{equation}
and the replicator dynamics becomes 
\begin{equation}
    \begin{split}
       & \dot{x}=x(1-x)( a_2y + a_3 ), \\
       & \dot{y}=y(1-y)(b_1x+b_3),
    \end{split}
    \label{replicator_DG}
\end{equation}
where $a_3=\beta_{14}-\beta_{24}$, and $b_3=\beta_{32}-\beta_{42}$.
Under the stated payoff-sampling model, $a_2b_1\ne0$ almost surely, and open-edge equilibria occur with probability zero. The candidates therefore consist almost surely of the four vertices and one interior candidate, denoted here by $A_5$. The latter is feasible only when both coordinates lie strictly in $(0,1)$. Their coordinates and Jacobian eigenvalues are:
\begin{itemize}
    \item at $A_1=(0,0)$: $\lambda_1=a_{3}$, $\lambda_2=b_{3}$.
    \item at $A_2=(0,1)$: $\lambda_1=a_2+a_3$, $\lambda_2=-b_3$.
    \item at $A_3=(1,0)$: $\lambda_1=-a_3$, $\lambda_2=b_1+b_3$.
    \item at $A_4=(1,1)$: $\lambda_1=-(a_2+a_3)$, $\lambda_2=-(b_1+b_3)$.
    \item at $A_5=(-\frac{b_3}{b_1},-\frac{a_3}{a_2})$, $\lambda_1=-\sqrt{\frac{a_3(a_2+a_3)b_3(b_1+b_3)}{a_2b_1}}$, $\lambda_2=\sqrt{\frac{a_3(a_2+a_3)b_3(b_1+b_3)}{a_2b_1}}$.
\end{itemize}
\begin{lemma}
\label{lem: 2}
Suppose that the active payoff entries in the two off-diagonal blocks are independent $\mathcal N(0,1)$ random variables, while the diagonal blocks are zero. Then:
\begin{enumerate}
    \item Each vertex $A_1,\ldots,A_4$ always exists and is locally asymptotically stable with probability $1/4$.
    \label{T4}
    \item The interior candidate $A_5$ exists with probability $1/4$. When it exists, it is either a centre or a saddle, and is never locally asymptotically stable.
    \label{T5}
\end{enumerate}
\end{lemma}
A centre has nearby closed trajectories and does not attract them. The Supplementary Material applies this structure to an ultimatum game, including stability of its continuous equilibrium set.

\subsection{Exclusive games}

Exclusive games retain only within-population payoffs: the off-diagonal blocks of $B$ are zero, giving $a_2=b_1=0$, $a_3=\beta_{12}-\beta_{22}$, and $b_3=\beta_{34}-\beta_{44}$:
\begin{equation}
\bordermatrix{
       & s_1        & s_2       & s_3    &   s_4 \cr
s_1     &  \beta_{11} & \beta_{12} & 0 & 0 \cr
s_2     & \beta_{21} & \beta_{22} & 0 & 0 \cr
s_3    & 0 & 0 & \beta_{33} & \beta_{34} \cr
s_4     & 0 & 0 & \beta_{43} & \beta_{44}
},
\end{equation}
and the dynamics is
\begin{equation}
    \begin{split}
       & \dot{x}=x(1-x)( a_1x + a_3 ), \\
       & \dot{y}=y(1-y)(b_2y+b_3).
    \end{split}
    \label{replicator_EG}
\end{equation}
Assume $a_1b_2\ne0$, as holds almost surely under the stated payoff-sampling model. The equilibrium candidates and their Jacobian eigenvalues are listed below. An open-edge or interior candidate is feasible only when all its free coordinates lie strictly in $(0,1)$.
\begin{itemize}
    \item at $A_1=(0,0)$: $\lambda_1=a_{3}$, $\lambda_2=b_{3}$.
    \item at $A_2=(0,1)$: $\lambda_1=a_3$, $\lambda_2=-b_2-b_3$.
    \item at $A_3=(1,0)$: $\lambda_1=-a_1-a_3$, $\lambda_2=b_3$.
    \item at $A_4=(1,1)$: $\lambda_1=-a_1-a_3$, $\lambda_2=-b_2-b_3$.
    \item at $A_5=(0,-\frac{b_3}{b_2})$: $\lambda_1=a_3$, $\lambda_2=-\frac{b_3(b_2+b_3)}{b_2}$.
    \item at $A_6=(1,-\frac{b_3}{b_2})$: $\lambda_1=-a_1-a_3$, $\lambda_2=-\frac{b_3(b_2+b_3)}{b_2}$.
    \item at $A_7=(-\frac{a_3}{a_1},0)$: $\lambda_1=-\frac{a_3(a_1+a_3)}{a_1}$, $\lambda_2=b_3$.
    \item at $A_8=(-\frac{a_3}{a_1},1)$: $\lambda_1=-\frac{a_3(a_1+a_3)}{a_1}$, $\lambda_2=-b_2-b_3$.
    \item at $A_9=(-\frac{a_3}{a_1},-\frac{b_3}{b_2})$, $\lambda_1=-\frac{a_3(a_1+a_3)}{a_1}$, $\lambda_2=-\frac{b_3(b_2+b_3)}{b_2}$.
\end{itemize}
\begin{lemma}
\label{lem: 4}
Suppose that the active payoff entries in the two diagonal blocks are independent $\mathcal N(0,1)$ random variables, while the off-diagonal blocks are zero. Then:
\begin{enumerate}
    \item Each vertex $A_1,\ldots,A_4$ always exists and is locally asymptotically stable with probability $1/4$.

    \item Each open-edge candidate $A_5,\ldots,A_8$ exists with probability $1/2$ and exists and is locally asymptotically stable with probability $1/8$.
    \label{item:exclusive:boundary}

    \item The interior candidate $A_9$ exists with probability $1/4$. It exists and is locally asymptotically stable with probability $1/16$, and exists and is a saddle with probability $1/8$.
\end{enumerate}
\end{lemma}

\subsection{Avoidance games}
Avoidance games retain only payoffs within population 1: all blocks of $B$ except the upper-left block are zero, leading to $a_2=b_1=b_2=b_3=0$ and $a_3=\beta_{12}-\beta_{22}$:
\begin{equation}
\bordermatrix{
       & s_1        & s_2       & s_3    &   s_4 \cr
s_1     &  \beta_{11} & \beta_{12} & 0 & 0 \cr
s_2     & \beta_{21} & \beta_{22} & 0 & 0 \cr
s_3    & 0 & 0 & 0 & 0 \cr
s_4     & 0 & 0 & 0 & 0
}.
\end{equation}
The resulting dynamics is
\begin{equation}
\begin{aligned}
\dot{x}&=x(1-x)(a_1x+a_3),\\
\dot{y}&=0.
\end{aligned}
\label{replicator_AG}
\end{equation}
Thus, population 2 retains its initial strategy frequency, $y(t)=y(0)$, while population 1 follows a scalar replicator equation. The scalar equilibrium candidates are
\[
A_1=0,\qquad A_2=1,\qquad A_3=-\frac{a_3}{a_1},
\]
where $A_3$ is admissible only when it lies in $(0,1)$.

Each admissible scalar equilibrium $A_i$ corresponds to a continuous family of equilibria in the full system:
\[
\{(A_i,y):0\leq y\leq1\}.
\]
If $A_i$ attracts nearby scalar states, trajectories converge to $(A_i, y(0))$. We call this transverse attraction: different initial $y$ coordinates lead to different limiting points, so no single point attracts a full two-dimensional neighbourhood.
\begin{lemma}
\label{lem: 5}
In the avoidance game, suppose that the four active payoff entries $\beta_{11},\beta_{12},\beta_{21},\beta_{22}$ are independent $\mathcal{N}(0,1)$ random variables. The following statements hold almost surely, with scalar local asymptotic stability understood relative to $[0,1]$ for the $x$ equation in \eqref{replicator_AG}.
\begin{enumerate}
    \item $A_1=0$ and $A_2=1$ always exist. They are stable if and only if $a_3<0$ and $a_1+a_3>0$, respectively, each with probability $1/2$.
    \item $A_3=-a_3/a_1$ exists in $(0,1)$ if and only if $a_3(a_1+a_3)<0$, with probability $1/2$. Given its existence, $A_3$ is stable if and only if $a_3>0$. Thus it exists and is stable if and only if $a_3>0>a_1+a_3$, with joint probability $1/4$; its conditional stability probability given existence is $1/2$.
    \item Scalar local asymptotic stability of $A_i$ is equivalent to transverse attraction of its corresponding equilibrium family. Every point $(A_i,y^*)$ in an attracting family is Lyapunov stable relative to $[0,1]^2$. No individual point $(A_i,y^*)$ is locally asymptotically stable in the full two-dimensional system, because it has arbitrarily close distinct equilibria in the same family.
\end{enumerate}
\end{lemma}

\section{Three-population analytical results}
\label{sec:3pop-analysis}

We next examine how the equilibrium statistics extend to three-population games. Here we return to the generic model, retain all within- and between-population payoff blocks, and consider independent payoff sampling. Related deterministic three-population models have been studied in \cite{zhang2022tripartite,zhao2020tripartite,zhou2022evolutionary}.

The state space is now a cube. Its vertices, open edges, open faces, and interior correspond to zero, one, two, and three populations maintaining both strategies, respectively. This classification extends the vertex--edge--interior description of the two-population state square. We first identify the candidates and their eigenvalues, and then give their existence and stability probabilities.

\begin{table*}[t!]
\caption{Theoretical and Empirical Equilibrium Probabilities in Generic Two-Population Games Under Microscopic Uncertainty}
\label{tab:table2}
\centering
\footnotesize
\renewcommand{\arraystretch}{1.2}
\begin{tabular*}{\textwidth}{@{\extracolsep{\fill}}llcccc@{}}
\toprule
\multicolumn{2}{c}{Equilibrium}
& \multicolumn{2}{c}{Existence: $\mathbb{P}(E_i)$}
& \multicolumn{2}{c}{Existence and stability: $\mathbb{P}(E_i\cap S_i)$} \\
\cmidrule(lr){1-2}\cmidrule(lr){3-4}\cmidrule(l){5-6}
Type & Point & Theory & Simulation & Theory & Simulation \\
\midrule
Vertex
& $A_1=(0,0)$ & $1$ & $1.0000$
& $1/4=0.2500$ & $0.2500$ \\
& $A_2=(0,1)$ & $1$ & $1.0000$
& $1/4=0.2500$ & $0.2497$ \\
& $A_3=(1,0)$ & $1$ & $1.0000$
& $1/4=0.2500$ & $0.2494$ \\
& $A_4=(1,1)$ & $1$ & $1.0000$
& $1/4=0.2500$ & $0.2510$ \\
\midrule
Open edge
& $A_5$ ($x=0$) & $1/3\approx0.3333$ & $0.3333$
& $1/12\approx0.0833$ & $0.0833$ \\
& $A_6$ ($x=1$) & $1/3\approx0.3333$ & $0.3332$
& $1/12\approx0.0833$ & $0.0834$ \\
& $A_7$ ($y=0$) & $1/3\approx0.3333$ & $0.3332$
& $1/12\approx0.0833$ & $0.0831$ \\
& $A_8$ ($y=1$) & $1/3\approx0.3333$ & $0.3328$
& $1/12\approx0.0833$ & $0.0835$ \\
\midrule
Interior
& $A_9=(x^*,y^*)$ & $p_{\mathrm{int}}\approx0.2163$
& $0.2163$ & $p_{\mathrm{int}}/4\approx0.0541$ & $0.0541$ \\
\bottomrule
\end{tabular*}

\vspace{4pt}
\begin{minipage}{\textwidth}
\footnotesize\raggedright
\textit{Notes:}
Analytical values follow from Theorem~\ref{thm: main-thm}.
\end{minipage}
\end{table*}

The full Jacobian is
\begin{equation}
\label{eq:3pop-Jacobian}
\begin{aligned}
J_{ij}(\boldsymbol u)
=\delta_{ij}(1-2u_i)f_i(\boldsymbol u)
+u_i(1-u_i)C_{ij},
\end{aligned}
\end{equation}
where $\delta_{ij}=1$ when $i=j$ and $0$ otherwise.
Equilibrium points and their eigenvalues are listed below, with $i,j,k\in\{0,1\}$. Every starred coordinate is determined separately for its candidate and must lie strictly in $(0,1)$. The formulas assume that the relevant principal submatrices of $C$ are nonsingular, as holds almost surely under independent Gaussian payoff sampling.
\begin{enumerate}[(i)]
\raggedright
\item \textit{Vertices.} The eight points $\boldsymbol u^*=(i,j,k)$ always exist. Their eigenvalues are
\[
\begin{aligned}
\lambda_1&=(1-2i)f_1(i,j,k),\\
\lambda_2&=(1-2j)f_2(i,j,k),\\
\lambda_3&=(1-2k)f_3(i,j,k).
\end{aligned}
\]
\item \textit{Open-edge points.} The twelve candidates have the forms $(x^*,i,j)$, $(i,y^*,j)$, and $(i,j,z^*)$. For $\boldsymbol u^*=(x^*,i,j)$,
\[
x^*=-\frac{a_2i+a_3j+a_4}{a_1},
\]
and the point exists on its open edge if and only if $0<x^*<1$. Its eigenvalues are
$\lambda_1=a_1x^*(1-x^*),$ 
$\lambda_2=(1-2i)f_2(x^*,i,j),$ and 
$\lambda_3=(1-2j)f_3(x^*,i,j)$.

The other two families follow by the corresponding permutations of coordinates and payoff-difference equations.
\item \textit{Open-face points.} The six candidates have the forms $(x^*,y^*,k)$, $(x^*,k,z^*)$, and $(k,y^*,z^*)$. For $\boldsymbol u^*=(x^*,y^*,k)$,
\begin{equation}
\begin{pmatrix}x^*\\y^*\end{pmatrix}
=-\begin{pmatrix}a_1&a_2\\b_1&b_2\end{pmatrix}^{-1}
\begin{pmatrix}a_3k+a_4\\b_3k+b_4\end{pmatrix}. 
\end{equation}
It exists on the open face if and only if $0<x^*,y^*<1$. 
With
\(
D_{xy}=\operatorname{diag}\bigl(x^*(1-x^*),y^*(1-y^*)\bigr) \) and 
\(K=D_{xy}\begin{pmatrix}a_1&a_2\\b_1&b_2\end{pmatrix},\)
the eigenvalues are
\[
\begin{aligned}
\lambda_{1,2}&=\frac{\operatorname{tr}K\pm
\sqrt{(\operatorname{tr}K)^2-4\det K}}{2},\\
\lambda_3&=(1-2k)f_3(x^*,y^*,k).
\end{aligned}
\]
The other two families again follow by corresponding permutations.
\item \textit{Interior point.} The candidate $\boldsymbol u^*=(x^*,y^*,z^*)=-C^{-1}\boldsymbol d$ exists if and only if $\boldsymbol u^*\in(0,1)^3$. Set $
D=\operatorname{diag}\bigl(u_r^*(1-u_r^*):r=1,2,3\bigr)$.
Its three eigenvalues are the roots, counted with multiplicity, of $
\det(\lambda I_3-DC)=0$.
\end{enumerate}
These four classes give 27 candidate locations and exhaust all equilibria under the stated nonsingularity assumption; only the eight vertices are guaranteed to exist. When no eigenvalue has zero real part, a candidate is locally asymptotically stable relative to $[0,1]^3$ if and only if all three eigenvalues have negative real parts. At a boundary equilibrium, this condition includes perturbations away from the boundary, not just perturbations along it. The corresponding derivations are provided in the Supplementary Material.

For a specified candidate with $k$ free coordinates, let $E$ denote existence and $S$ denote local asymptotic stability relative to $[0,1]^3$. Write $p_k=\mathbb P(E)$ and $q_k=\mathbb P(E\cap S)$ for the per-candidate probabilities; $k=0,1,2,3$ corresponds to a vertex, open-edge point, open-face point, or interior point, respectively.

\begin{table*}[t!]
\caption{Empirical Equilibrium-Count Distributions in Generic Two-Population Games Under Microscopic Uncertainty}
\label{tab:table4}
\centering
\footnotesize
\renewcommand{\arraystretch}{1.2}
\begin{tabular*}{\textwidth}{@{\extracolsep{\fill}}l*{10}{c}@{}}
\toprule
Empirical probability
& \multicolumn{10}{c}{Number of equilibria, $n$} \\
\cmidrule(l){2-11}
& $0$ & $1$ & $2$ & $3$ & $4$
& $5$ & $6$ & $7$ & $8$ & $9$ \\
\midrule
$\widehat{\mathbb{P}}(N=n)$
& $0$ & $0$ & $0$ & $0$
& $0.2222$ & $0.3103$ & $0.2703$
& $0.1184$ & $0.0511$ & $0.0277$ \\
\midrule
$\widehat{\mathbb{P}}(N_{\mathrm{s}}=n)$
& $0.0074$ & $0.6465$ & $0.3044$ & $0.0349$
& $0.0069$ & $0$ & $0$ & $0$ & $0$ & $0$ \\
\bottomrule
\end{tabular*}

\vspace{4pt}
\begin{minipage}{\textwidth}
\footnotesize\raggedright
\textit{Notes:}
Both marginal distributions use the same payoff draws as Table~\ref{tab:table2}. Zero denotes no observed occurrences; $N<4$ is impossible because all four vertices always exist. Rounding may cause row sums to differ slightly from one.
\end{minipage}
\end{table*}

\begin{figure*}[t!]
\centering
\includegraphics[width=0.7\textwidth]{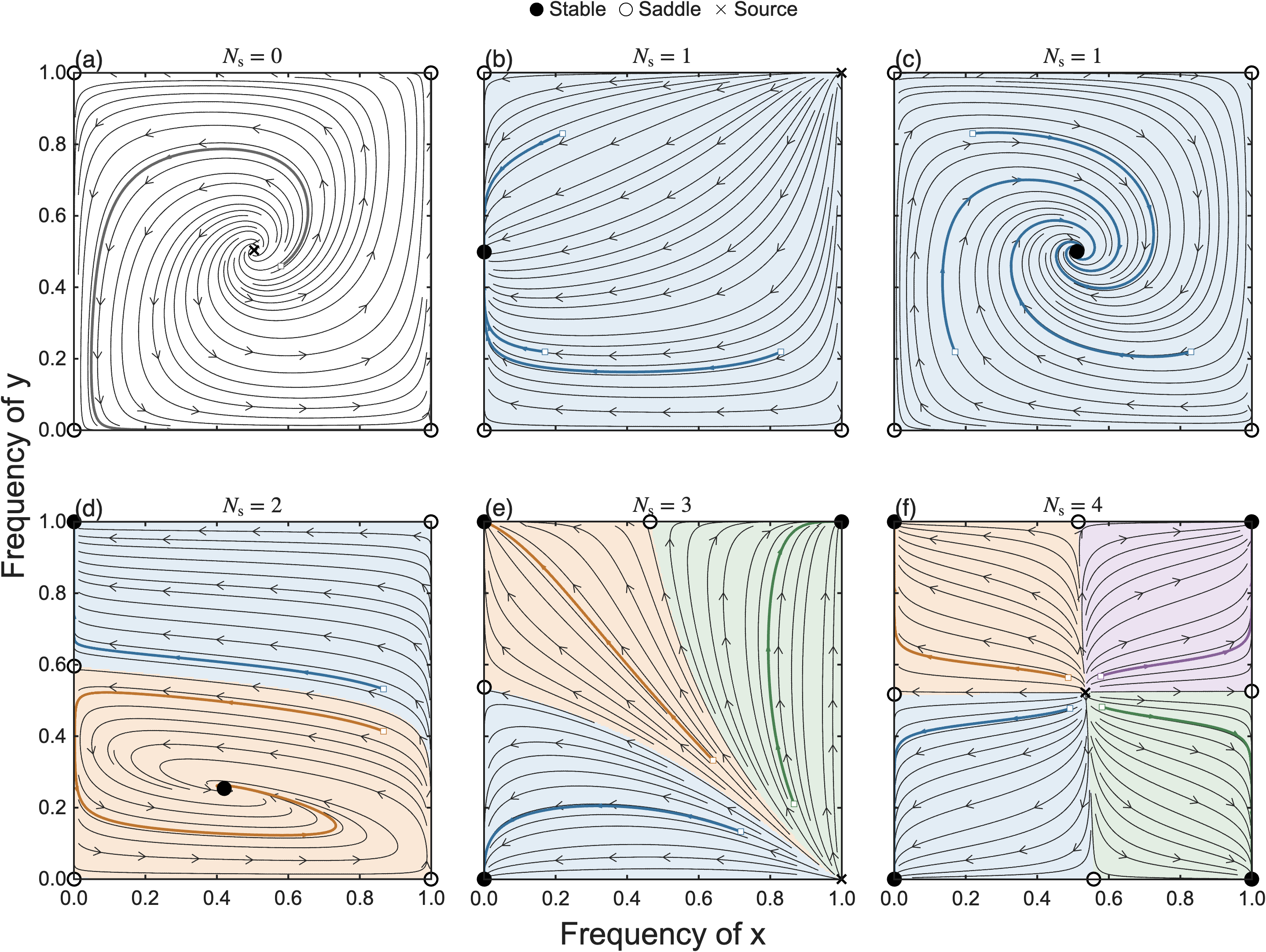}
\caption[Representative phase portraits under microscopic uncertainty]{Representative phase portraits and initial-condition dependence. Panels (a)--(f) contain $N_{\mathrm{s}}=0,1,1,2,3,4$ locally asymptotically stable equilibria, respectively. Panel (a) has an attracting boundary heteroclinic cycle, not a stable equilibrium or an interior limit cycle. The unique stable equilibrium is on the boundary in (b) and in the interior in (c); (d) exhibits coexistence of a stable vertex and a stable interior equilibrium. Black filled circles denote stable equilibria, open circles denote saddles, and crosses denote unstable sources. No neutrally stable equilibria occur in these samples. Black arrows indicate the vector field. Pale colours in (b)--(f) identify numerically estimated basins of attraction within each panel; darker curves illustrate trajectories from selected initial states, marked by small squares. The examples were selected from payoff-sampled games to illustrate distinct dynamical regimes; their selection does not represent the relative frequencies of those regimes. Writing $\boldsymbol{\theta}=(a_1,a_2,a_3,b_1,b_2,b_3)$, the parameter sets are:
(a) $\boldsymbol{\theta}=(1.5632,\allowbreak -2.2136,\allowbreak 0.3277,\allowbreak 2.2314,\allowbreak 0.6897,\allowbreak -1.4731)$;
(b) $\boldsymbol{\theta}=(1.5693,\allowbreak 2.2535,\allowbreak -6.6246,\allowbreak -3.6760,\allowbreak -4.3716,\allowbreak 2.1825)$;
(c) $\boldsymbol{\theta}=(-0.9148,\allowbreak 4.0690,\allowbreak -1.5678,\allowbreak -2.8839,\allowbreak -2.0114,\allowbreak 2.4812)$;
(d) $\boldsymbol{\theta}=(-2.1943,\allowbreak -6.5550,\allowbreak 2.5883,\allowbreak 0.5775,\allowbreak 0.7098,\allowbreak -0.4233)$;
(e) $\boldsymbol{\theta}=(2.3992,\allowbreak 2.2562,\allowbreak -3.3708,\allowbreak 2.6826,\allowbreak 2.5720,\allowbreak -1.3808)$;
(f) $\boldsymbol{\theta}=(4.6340,\allowbreak 0.2002,\allowbreak -2.5835,\allowbreak -0.0118,\allowbreak 1.5367,\allowbreak -0.7956)$. The listed parameter values are rounded to four decimal places.}
\label{fig:beta-phase-portraits}
\end{figure*}
\begin{lemma}[Probabilities under microscopic uncertainty]
\label{lem:3pop}
Suppose that all 36 payoff entries $\beta_{ij}$ are independent $\mathcal N(0,1)$ variables. For each specified candidate, the following statements hold.
\begin{enumerate}[(i)]
\item Each vertex always exists and is locally asymptotically stable with probability $1/8$:
\[
p_0=1,\qquad q_0=\frac18.
\]
\item Each open-edge point exists with probability $p_1$ and exists and is locally asymptotically stable with probability $q_1$, where
\[
\begin{aligned}
p_1&=\frac{1}{\pi}\arccos\!\left(\frac23\right)\approx0.2677,\\
q_1&=\frac{1}{8\pi}\arccos\!\left(\frac23\right)\approx0.0335.
\end{aligned}
\]
\item Each open-face point exists with probability $p_2$ and exists and is locally asymptotically stable with probability $q_2$, where
\[
\begin{aligned}
p_2&=\frac{2}{\pi}\arctan\!\left(\frac{1}{2\sqrt6}\right)\approx0.1282,\\
q_2&=\frac{1}{4\pi}\arctan\!\left(\frac{1}{2\sqrt6}\right)\approx0.0160.
\end{aligned}
\]
\item The interior point exists with probability
\begin{equation}
\label{eq:3pop-interior-probability}
p_3=2\int_0^\infty\phi(t)
       \left[2\Phi\!\left(\frac{t}{\sqrt3}\right)-1\right]^3\,dt
       \approx0.0922,
\end{equation}
where $\phi$ and $\Phi$ are the standard normal density and distribution function.
\end{enumerate}
\end{lemma}

The proof is given in the Supplementary Material. The probabilities $q_0,q_1,q_2$ are joint, not conditional, probabilities; the corresponding conditional stability probabilities equal $1/8$. The lemma does not give an analytical value for the interior-stability probability $q_3$, which is assessed numerically.

\section{Numerical results}
\label{sec: numerics}
\begin{table*}[htp]
\caption{Effect of Interaction Structure on Equilibrium Existence and Stability}
\label{tab:interaction-structure}
\centering\footnotesize
\renewcommand{\arraystretch}{1.2}
\begin{tabular*}{\textwidth}{@{\extracolsep{\fill}}llcccc@{}}
\toprule
Model & Entity type & \multicolumn{2}{c}{Existence: $\mathbb P(E)$} & \multicolumn{2}{c}{Joint: $\mathbb P(E\cap S)$ or $\mathbb P(E\cap T)$} \\
\cmidrule(lr){3-4}\cmidrule(l){5-6}
 & & Theory & Simulation & Theory & Simulation \\
\midrule
\multicolumn{6}{l}{\textit{(a) Isolated equilibria: $S$ denotes local asymptotic stability}} \\
Generic & Vertices & $1$ & 1.0000 & $1/4$ & 0.2499 \\
 & Open-edge points & $1/3$ & 0.3331 & $1/12$ & 0.0833 \\
 & Interior point & $p_{\mathrm{int}}$ & 0.2160 & $p_{\mathrm{int}}/4$ & 0.0539 \\
\addlinespace[2pt]
Diversity & Vertices & $1$ & 1.0000 & $1/4$ & 0.2501 \\
 & Open-edge points & $0$ & 0.0000 & $0$ & 0.0000 \\
 & Interior point & $1/4$ & 0.2497 & $0$ & 0.0000 \\
\addlinespace[2pt]
Exclusive & Vertices & $1$ & 1.0000 & $1/4$ & 0.2500 \\
 & Open-edge points & $1/2$ & 0.5007 & $1/8$ & 0.1251 \\
 & Interior point & $1/4$ & 0.2508 & $1/16$ & 0.0629 \\
\midrule
\multicolumn{6}{l}{\textit{(b) Avoidance: $T$ denotes transverse attraction of the corresponding equilibrium family}} \\
Avoidance & Families at $x=A_1,A_2$
& $1$ & 1.0000 & $1/2$ & 0.5001 \\
 & Family at $x=A_3$
& $1/2$ & 0.5010 & $1/4$ & 0.2504 \\
\bottomrule
\end{tabular*}
\vspace{4pt}

\begin{minipage}{\textwidth}\footnotesize\raggedright
\textit{Notes:} In (b), $A_i$ denotes a scalar equilibrium from Lemma~\ref{lem: 5}; each corresponding family is $\{(A_i,y):0\le y\le1\}$. For $A_3\in(0,1)$, the family includes boundary endpoints at $y=0,1$. The generic-game estimates use a separate realization set from Table~\ref{tab:table2}; $p_{\mathrm{int}}$ is defined in Theorem~\ref{thm: main-thm}(2).
\end{minipage}
\end{table*}

\begin{table*}[htp]
\caption{Theoretical and Empirical Equilibrium Probabilities in Generic Three-Population Games Under Microscopic Uncertainty}
\label{tab:3pop-beta-probabilities}
\centering\footnotesize
\renewcommand{\arraystretch}{1.15}
\setlength{\tabcolsep}{4pt}
\begin{tabular*}{\textwidth}{@{\extracolsep{\fill}}lrcccc@{}}
\toprule
& & \multicolumn{2}{c}{Existence} & \multicolumn{2}{c}{Existence and stability} \\
\cmidrule(lr){3-4}\cmidrule(l){5-6}
Type & Candidates & Theory & Simulation & Theory & Simulation \\
\midrule
Vertex & 8 & $1.0000$ & $1.0000\pm0.0000$ & $0.1250$ & $0.1251\pm0.0004$ \\
Open edge & 12 & $0.2677$ & $0.2678\pm0.0003$ & $0.0335$ & $0.0335\pm0.0002$ \\
Open face & 6 & $0.1282$ & $0.1284\pm0.0004$ & $0.0160$ & $0.0161\pm0.0001$ \\
Interior & 1 & $0.0922$ & $0.0923$ & --- & $0.0096$ \\
\bottomrule
\end{tabular*}
\vspace{4pt}

\begin{minipage}{\textwidth}\footnotesize\raggedright
\textit{Notes:} Theory is given by Lemma~\ref{lem:3pop}; the interior-existence value evaluates its exact integral. Simulation entries are mean $\pm$ between-candidate SD, omitted for the single interior candidate. The dash denotes an analytical interior-stability probability not established here.
\end{minipage}
\end{table*}

The numerical results support the analytical distinction between equilibrium existence and attraction, and show how it relates to multistability and initial-condition dependence. The comparison covers generic two-population games, the three interaction structures, and generic three-population games. Additional coefficient-sampling results and model-specific dynamical examples are provided in the Supplementary Material.

\subsection{Generic two-population games}

Equilibrium statistics depend on whether uncertainty is specified at the payoff or coefficient level. Under independent payoff sampling, the interior equilibrium exists with frequency $0.2163$ in Table~\ref{tab:table2}, consistent with the analytical value $p_{\mathrm{int}}\approx0.21635$. Independent coefficient sampling instead gives $1/12\approx0.0833$, supported by the supplementary simulations. Likewise, the four open-edge candidates share existence probability $1/3$ under payoff sampling, whereas coefficient sampling produces two pairs with distinct probabilities. Thus, the two uncertainty descriptions define different evolutionary benchmarks. 

Under payoff sampling, a specified equilibrium is locally asymptotically stable with probability $1/4$ conditional on its existence, regardless of whether it lies at a vertex, on an open edge, or in the interior. The joint existence-and-stability probabilities are therefore $1/4$, $1/12$, and $p_{\mathrm{int}}/4\approx0.0541$, respectively, consistent with Table~\ref{tab:table2}. Equilibrium existence is consequently substantially more common than attraction.

Multistability occurs in more than one third of payoff-sampled games, although a single stable equilibrium is the most common outcome. This concerns the coexistence of attractors within the same game, which is not determined by the probabilities for individual candidates alone. In Table~\ref{tab:table4}, $34.62\%$ of games have at least two locally asymptotically stable equilibria, $64.65\%$ have exactly one, and $0.74\%$ have none. The mean counts are $5.5489$ existing equilibria and $1.3874$ stable equilibria. Five equilibria is the most frequent total count ($31.03\%$), illustrating that most existing equilibria need not be attractors.

Coexisting stable equilibria make the long-term outcome depend on the initial strategy frequencies even when payoffs are fixed. Panels (d)--(f) of Fig.~\ref{fig:beta-phase-portraits} display two, three, and four stable equilibria, respectively. Each attracts a different set of initial states, forming its basin of attraction. In particular, panel (d) admits both a stable vertex and a stable interior equilibrium, so initial conditions can select between pure-strategy and mixed-strategy outcomes. Panels (b) and (c) instead have a unique stable equilibrium on the boundary and in the interior, respectively. 

The absence of a stable equilibrium does not preclude an attracting dynamical outcome. In Fig.~\ref{fig:beta-phase-portraits}(a), trajectories approach a boundary heteroclinic cycle: they repeatedly pass near successive boundary saddle points rather than converge to a single equilibrium. The interior equilibrium is unstable, so this recurrent switching is distinct from attraction to a stable coexistence point. 

\begin{table*}[t!]
\caption{Empirical Equilibrium-Count Distributions in Generic Three-Population Games Under Microscopic Uncertainty}
\label{tab:3pop-beta-counts}
\centering\footnotesize
\renewcommand{\arraystretch}{1.2}
\setlength{\tabcolsep}{3pt}
\begin{tabular*}{\textwidth}{@{\extracolsep{\fill}}l*{10}{c}@{}}
\toprule
\multicolumn{11}{l}{\textit{(a) Total number of equilibria}} \\
$n$ & $8$ & $9$ & $10$ & $11$ & $12$ & $13$ & $14$ & $15$ & $16$ & $17$ \\
\cmidrule(l){2-11}
$\widehat{\mathbb P}(N=n)$ & 0.0582 & 0.1033 & 0.1543 & 0.1569 & 0.1499 & 0.1173 & 0.0873 & 0.0634 & 0.0412 & 0.0263 \\
\addlinespace[3pt]
$n$ & $18$ & $19$ & $20$ & $21$ & $22$ & $23$ & $24$ & $25$ & $26$ & $27$ \\
\cmidrule(l){2-11}
$\widehat{\mathbb P}(N=n)$ & 0.0178 & 0.0098 & 0.0059 & 0.0037 & 0.0019 & 0.0013 & 0.0007 & 0.0003 & 0.0002 & 0.0001 \\
\midrule
\multicolumn{11}{l}{\textit{(b) Number of locally asymptotically stable equilibria}} \\
$n$ & $0$ & $1$ & $2$ & $3$ & $4$ & $5$ & $6$ & $7$ & $8$ & $9$--$27$ \\
\cmidrule(l){2-11}
$\widehat{\mathbb P}(N_{\mathrm s}=n)$ & 0.0130 & 0.5728 & 0.3285 & 0.0674 & 0.0156 & 0.0021 & 0.0005 & 0.0001 & $<0.0001$ & 0 \\
\bottomrule
\end{tabular*}
\vspace{4pt}

\begin{minipage}{\textwidth}\footnotesize\raggedright
\textit{Notes:} The same experiment underlies Table~\ref{tab:3pop-beta-probabilities}. Panel (a) splits one distribution across two ranges. The last entry of (b) aggregates $9\leq N_{\mathrm s}\leq27$; the positive entry $<0.0001$ corresponds to 17 games.
\end{minipage}
\end{table*}

\subsection{Effects of interaction structure}
Interaction structure also changes the availability of partially mixed outcomes, in which one population maintains both strategies while the other uses a pure strategy. For each specified open edge, the equilibrium-existence probability is zero in diversity games, $1/3$ in generic games, and $1/2$ in exclusive games. The corresponding joint existence-and-stability probabilities are zero, $1/12$, and $1/8$, consistent with Table~\ref{tab:interaction-structure}.

With the payoff-sampling assumption held fixed, interaction structure changes stability even when the probability of interior coexistence remains the same. Diversity and exclusive games both have interior-existence probability $1/4$, supported by the frequencies $0.2497$ and $0.2508$ in Table~\ref{tab:interaction-structure}. Yet diversity games admit no asymptotically stable interior equilibrium, whereas exclusive games have joint interior-stability probability $1/16$, with simulated frequency $0.0629$. Interior coexistence alone therefore does not identify an attracting mixed-strategy outcome. Detailed candidate-wise probabilities and equilibrium-count
distributions are reported in Tables S3 and S4 of the
Supplementary Material.

Avoidance changes the attracting object from an isolated equilibrium to a continuous family. The family associated with the interior scalar equilibrium $A_3\in(0,1)$ exists with frequency $0.5010$ and exists and is transversely attracting with frequency $0.2504$, consistent with the probabilities $1/2$ and $1/4$ in Lemma~\ref{lem: 5}. Boundary families have mean transverse-attraction frequency $0.5001$. None of these families contains an individually locally asymptotically stable point of the full two-dimensional system. Detailed candidate-wise results and equilibrium-count distributions are provided in the Supplementary Material.

\subsection{Three-population numerical results}
\label{sec:3-population}
Within the three-population model, a specified equilibrium candidate is less likely to exist when it requires more populations to maintain both strategies. In Table~\ref{tab:3pop-beta-probabilities}, the per-candidate existence frequencies decrease from $1$ at vertices to $0.2678$ on open edges, $0.1284$ on open faces, and $0.0923$ in the interior, consistent with Lemma~\ref{lem:3pop}. Interior coexistence is substantially more frequent than stable interior coexistence: the joint interior-stability frequency is only $0.0096$. For each boundary candidate, the analytical joint stability probability is one eighth of its existence probability. Equivalently, its conditional probability of local asymptotic stability, given existence, is $1/8$. No corresponding analytical relation is established here for the interior candidate.

Under the independent Gaussian payoff model with equal interaction weights, multistability is more frequent in the three-population sample than in the two-population sample, despite the lower probability of a fully mixed equilibrium. Games with at least two locally asymptotically stable equilibria account for $41.42\%$ of Table~\ref{tab:3pop-beta-counts}, compared with $34.62\%$ in Table~\ref{tab:table4}. A single stable equilibrium remains the most common outcome ($57.28\%$), whereas $1.30\%$ of games have none. 
These count distributions characterise multistability, not global convergence: games without a stable equilibrium may still possess non-equilibrium attractors. Additional three-dimensional trajectories are provided in the Supplementary Material.

\section{Conclusion}
\label{conclusion}
We establish probabilistic benchmarks for equilibrium existence and stability in two-population games, with a generic three-population extension. Our results show that equilibrium statistics depend both on interaction structure and on how uncertainty is represented. In particular, independent sampling of payoff entries differs from independent sampling of replicator coefficients, because shared payoff entries induce dependence among the coefficients.
The structured two-population games illustrate why equilibrium existence and attraction must be distinguished. Diversity and exclusive games have the same probability of an interior equilibrium, but differ in whether stable interior coexistence is possible. In avoidance games, equilibria may form continuous families that attract trajectories transversely, even though individual points on those families are not asymptotically stable. This distinction is essential to stability analyses of asymmetric games, where equilibrium sets need not be isolated~\cite{song2024study,song2024stability}.

Under the Gaussian payoff ensembles considered here, fully mixed equilibria occur less frequently in three-population games than in two-population games, whereas multistability is more frequent. These findings are benchmarks for the specified uncertainty models, rather than universal properties of evolutionary games. Payoffs remain fixed during each trajectory; thus, the probabilities describe variation across games, not stochastic forcing or changes in the game during evolution, which are studied separately~\cite{hong2026multi,feng2022stochastic,feng2023evolutionary}. Future work could investigate how the results change under correlated or empirically calibrated payoffs, and develop analytical stability results for interior equilibria in three-population games.

\section*{Acknowledgment}
Z. Song and T.A.H. were supported by EPSRC (grant no. EP/Y00857X/1). M.H.D is
supported by EPSRC (grant no. EP/Y008561/1).
\bibliographystyle{IEEEtran}
\bibliography{mybib} 
\end{document}